\documentclass[aps,prl,reprint,superscriptaddress]{revtex4-1}
\usepackage{graphicx,xcolor,bm,amsmath,amssymb,natbib}

\newcommand{\UPt}{UPt$_3$}

\begin{document}

\title{Effects of the order parameter anisotropy on the vortex lattice in {\UPt}}

\author{K.~E.~Avers}
\altaffiliation[Current address: ]{Department of Physics, University of Maryland, College Park, Maryland 20742, USA}
\affiliation{Department of Physics and Astronomy, Northwestern University, Evanston, Illinois 60208, USA}
\affiliation{Center for Applied Physics \& Superconducting Technologies, Northwestern University, Evanston, Illinois 60208, USA}
%\affiliation{Los Alamos National Laboratory, Los Alamos, NM 87545, USA}

\author{W.~J.~Gannon}
\affiliation{Department of Physics and Astronomy, Northwestern University, Evanston, Illinois 60208, USA}
\affiliation{Department of Physics and Astronomy, University of Kentucky, Lexington, Kentucky 40506, USA}

\author{A.~W.~D.~Leishman}
\affiliation{Department of Physics, University of Notre Dame, Notre Dame, Indiana 46556, USA}

\author{L.~DeBeer-Schmitt}
\affiliation{Large Scale Structures Section, Neutron Scattering Division, Oak Ridge National Laboratory, Oak Ridge, Tennessee 37831, USA}

\author{W.~P.~Halperin}
\affiliation{Department of Physics and Astronomy, Northwestern University, Evanston, Illinois 60208, USA}

\author{M.~R.~Eskildsen}
\email[email: ]{eskildsen@nd.edu}
\affiliation{Department of Physics, University of Notre Dame, Notre Dame, Indiana 46556, USA}

\date{\today}

%%% ABSTRACT
\begin{abstract}
We have used small-angle neutron scattering to determine the vortex lattice phase diagram in the topological superconductor {\UPt} for the applied magnetic field along the crystalline $c$-axis.
A triangular vortex lattice is observed throughout the superconducting state, but with an orientation relative to the hexagonal basal plane  that changes with field and temperature.
At low temperature, in the chiral B phase, the vortex lattice undergoes a non-monotonic rotation with increasing magnetic field.
The rotation amplitude decreases with increasing temperature and vanishes before reaching the A phase.
Within the A phase an abrupt $\pm 15^{\circ}$ vortex lattice rotation was previously reported by Huxley {\em et al.}, Nature {\bf 406}, 160-164 (2000).
The complex phase diagram may be understood from competing effects of the superconducting order parameter, the symmetry breaking field, and the Fermi surface anisotropy.
The low-temperature rotated phase, centered around 0.8~T, reported by Avers {\em et al.}, Nature Physics {\bf 16}, 531-535 (2020), can be attributed directly to the symmetry breaking field.
\end{abstract}

\maketitle

%%%%% INTRODUCTION %%%%%
\section{Introduction}
With three distinct superconducting phases {\UPt} has attracted significant attention~\cite{Joynt:2002wt}, but despite decades of experimental and theoretical studies the unconventional superconductivity in this material is still not fully understood.
Figure~\ref{VLrotation}(b) shows the {\UPt} phase diagram, indicting the extent of the superconducting A, B and C phases.
The presence of two distinct zero-field superconducting transitions suggests that the order parameter belongs to one of the two-dimensional representations of the D$_{6h}$ point group~\cite{Hess:1989wt}.
Here, $f$-wave pairing states with the $E_{2u}$ irreducible representation are the most likely~\cite{Sauls:1994wd}.
In such a scenario the B phase breaks time reversal and mirror symmetries while the A and C phases are time-reversal symmetric.
Experimental support comes from the $H$-$T$ phase diagram~\cite{Adenwalla:1990we,Sauls:1994wd,Shivaram:1986wo,Choi:1991wa}, and thermodynamic and transport studies~\cite{Taillefer:1997eq,Graf:2000ua}.
Broken time-reversal symmetry in the B phase is supported by phase-sensitive Josephson tunneling~\cite{Strand:2009eq}, the observation of polar Kerr rotation~\cite{Schemm:2014fv}, and a field history-dependent vortex lattice (VL) configuration~\cite{Avers:2020wx}.
Finally, the linear temperature dependence of the London penetration depth is consistent with a quadratic dispersion of the energy gap at the polar nodes structure, which is a characteristic of the $E_{2u}$ model~\cite{Signore:1995hu,Schottl:1999ks,Gannon:2015ct}.

A key component in the understanding of superconductivity in {\UPt} is the presence of a symmetry breaking field (SBF) that couples to the $E_{2u}$ superconducting order parameter~\cite{Hayden:1992gp}.
The SBF lifts the degeneracy of the multi-dimensional representation, splitting the zero-field transition and leading to the multiple superconducting phases~\cite{Sauls:1994wd}.
However, the origin of the SBF is an outstanding issue, with possible candidates that include a quasi-static antiferromagnetic state that develops at 5~K above the superconducting transition~\cite{Aeppli:1988ur,Aeppli:1988gw,Hayden:1992gp}, a distortion of the hexagonal crystal structure~\cite{Walko:2001fm}, or prismatic plane stacking faults~\cite{Hong:1999ud,Gannon:2012fu}.

Vortices provide a highly sensitive probe of the host superconductor.
This includes anisotropies in the screening current plane perpendicular to the applied magnetic field which affect the VL symmetry and orientation.
Such anisotropies may arise from the Fermi surface~\cite{Kogan:1981vl,Kogan:1997vm}, and nodes in or distortions of the superconducting gap~\cite{Huxley:2000aa,Avers:2020wx}.
As an example one can consider the ``simple'' superconductor niobium that displays a rich VL phase diagram when the applied field is along the [100] crystalline direction and the Fermi surface anisotropy is incommensurate with an equilateral triangular VL~\cite{Laver:2006bn,Laver:2009bk,Muhlbauer:2009hw}.
Even in materials with a hexagonal crystal structure VL rotations may occur due to competing anisotropies, as observed in MgB$_2$ when the applied field is perpendicular to the basal plane~\cite{Cubitt:2003ab,Das:2012cf}.

We have used small-angle neutron scattering (SANS) to determine the VL phase diagram in {\UPt}.
This extends our previous studies at low temperature, where the VL was found to undergo a field-driven, non-monotonic rotation transition~\cite{Avers:2020wx}.
We discuss how the VL phase diagram and the existence of the VL rotation transition can be directly attributed to the SBF.

\begin{figure*}
    \includegraphics[width = 168mm]{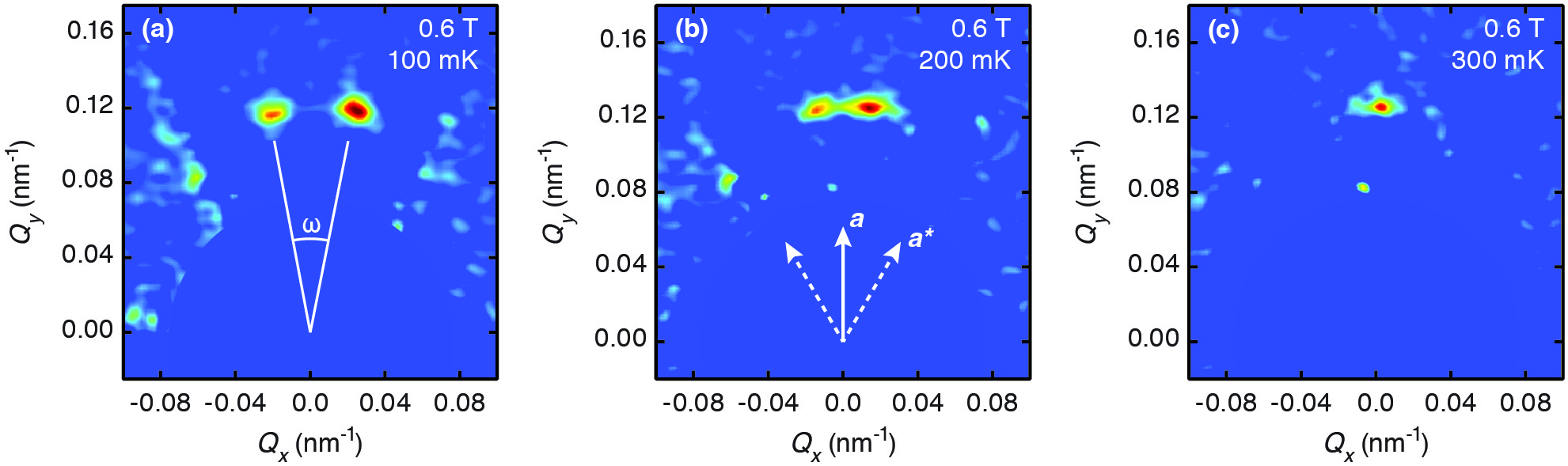}
    \caption{\label{DifPat}
        SANS VL diffraction patterns obtained at $H = 0.6$~T and $T = 100$~mK (a), 200~mK (b) and 300~mK (c).
        The Bragg peak splitting ($\omega$) is indicated in (a) and crystallographic directions within the scattering plane in (b).
        Only peaks at the top of the detector were imaged.
        Zero field background scattering is subtracted, and the detector center near $Q = 0$ is masked off.
        %\red{It can clearly be observed that while two Bragg peaks are resolvable at 100~mK and 200~mK , they merge into a single azimuthaly broadened peak within instrument resolution at 300~mK.}
        }
\end{figure*}
%

%%%%% EXP DETAILS %%%%%
\section{Experimental Details}
% All data at 100 mK and above was obtained on ZR11 at ORNL during the summer 2018 experiment.
Small-angle neutron scattering studies of the VL are possible due to the periodic field modulation from the vortices~\cite{Muhlbauer:2019jt}.
The scattered intensity depends strongly on the superconducting penetration depth, and for {\UPt} with a large in-plane $\lambda_{ab} \sim 680$~nm~\cite{Gannon:2015ct} necessitates a large sample volume.
For this work we used a high-quality single crystal (ZR11), combined with previously published results obtained on a separate sample (ZR8)~\cite{Avers:2020wx}.
Properties of both single crystals are listed in table~\ref{Xtals},
\begin{table}[b]
    \begin{tabular}{lcccc}
    \hline \hline
    Sample \hspace{0.1cm} & \hspace{0.1cm} mass~(g) \hspace{0.1cm} & \hspace{0.1cm} RRR \hspace{0.1cm} & \hspace{0.1cm} $T_{\rm c}$~(mK) \hspace{0.1cm} & \hspace{0.1cm} $\Delta T_{\rm c}$~(mK) \\ \hline
    ZR8  & 15 & $> 600$ & $560 \pm 2$ & 10 \\
    ZR11 &  9 & $> 900$ & $557 \pm 2$ &  5 \\ \hline \hline
    \end{tabular}
    \caption{\label{Xtals}
        Properties of the two {\UPt} single crystals used for the SANS experiments.}
\end{table}
determined from resistive measurements performed on smaller samples cut from the main crystals.
Here, RRR is the residual resistivity ratio, $T_{\rm c}$ is the superconducting transition temperature and $\Delta T_{\rm c}$ is the width of the transition.

For the SANS measurements each long, rod-like crystal was cut into two pieces, co-aligned and fixed with silver epoxy (EPOTEK E4110) to a copper cold finger.
The sample assembly was mounted onto the mixing chamber of a dilution refrigerator and placed inside a superconducting magnet, oriented with the crystalline $\textbf{a}$ axis vertical and the $\textbf{c}$ axis horizontally along the magnetic field and the neutron beam.
The neutron beam was masked off to illuminate a $7 \times 11$~mm$^2$ area.

The SANS experiment was performed at the GP-SANS beam line at the High Flux Isotope Reactor at Oak Ridge National Laboratory~\cite{Heller:2018gq}.
All measurements were carried out in a ``rocked on'' configuration, satisfying the Bragg condition for VL peaks at the top of the two-dimensional position sensitive detector, as seen in Fig.~\ref{DifPat}.
Background measurements, obtained either in zero field or above $H_{\text{c2}}$, were subtracted from both the field reduction and field reversal data.

Measurements were performed at temperatures between 100~mK and 300~mK and fields between 0.4~T and 1.2~T.
Prior to the SANS measurements the field was reduced from above the B-C phase transition at base temperature.
The sample was then heated to the measurement temperature and a damped field oscillation with an initial amplitude of 20~mT was applied to obtain a well ordered VL with a homogeneous vortex density~\cite{Avers:2020wx}.
Furthermore, a 5~mT field oscillation was applied approximately every 60 seconds during the SANS measurements, in order to counteract VL disordering due to neutron induced fission of $^{235}$U~\cite{Avers:2021wu}.

%%%%% RESULTS %%%%%
\section{Results}
Figure~\ref{DifPat} shows VL diffraction patterns obtained in an applied field of 0.6~T and temperature between 100~mK and 300~mK.
As previously reported, the VL in {\UPt} has a triangular symmetry but is in general not oriented along a high symmetry direction of the hexagonal crystalline basal lattice ($\bm{a}$ or $\bm{a}^*$)~\cite{Avers:2020wx}.
This causes the VL to break up into clockwise and counterclockwise rotated domains, and gives rise to the Bragg peak splitting in Figs.~\ref{DifPat}(a) and \ref{DifPat}(b).
With increasing temperature the splitting decreases, and the two peaks eventually merge as seen in Fig.~\ref{DifPat}(c).

\begin{figure}
    \includegraphics[width = 66mm]{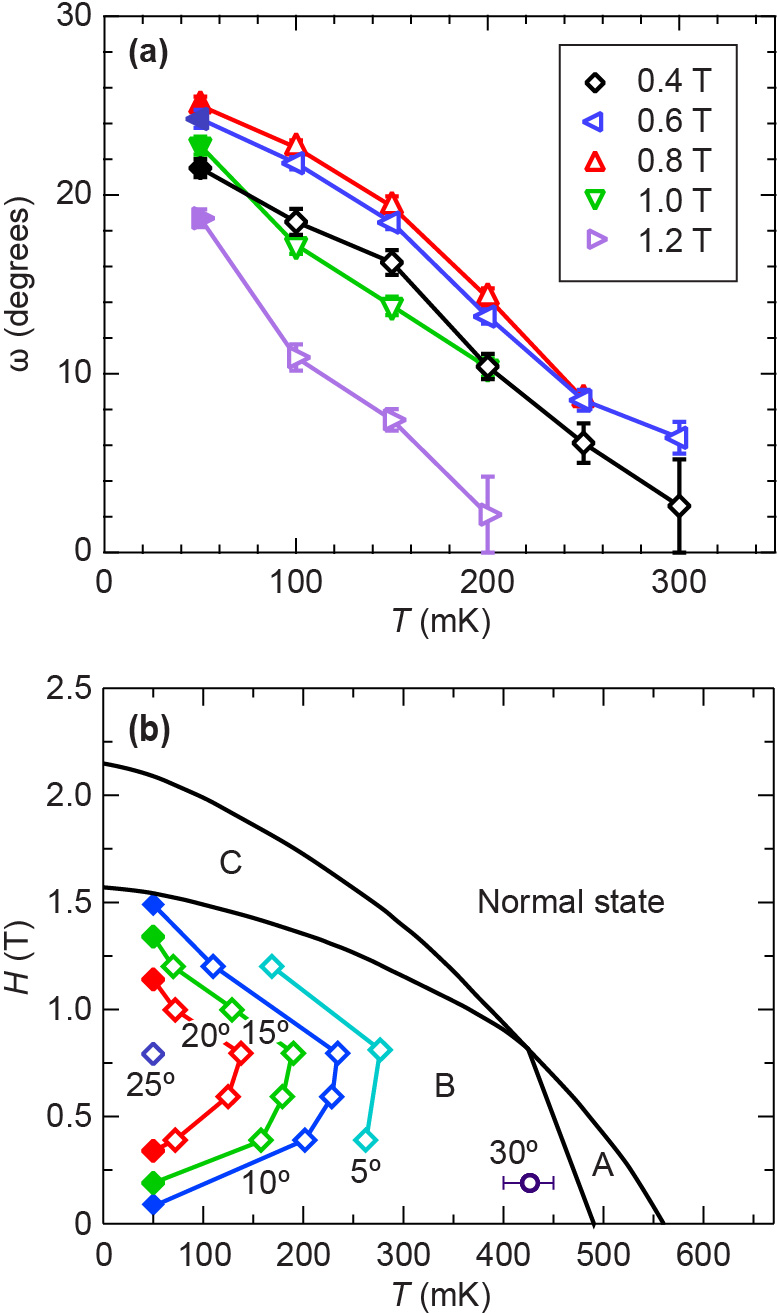}
    \caption{\label{VLrotation}
        Vortex lattice rotation.
        (a) VL peak splitting vs temperature for different magnetic fields.
        The data at 50~mK (solid symbols) was previously obtained on the ZR8 crystal~\cite{Avers:2020wx}.
        Error bars represent one standard deviation.
        (b) Constant $\omega$ contours superimposed on the {\UPt} phase diagram.
        Values are obtained from the data in (a) by interpolation (open diamonds) and from the 50~mK field dependence in Ref.~\onlinecite{Avers:2020wx} (solid diamonds).
        The $30^{\circ}$ data point (open circle) is from previous work by Huxley {\em et al.}~\cite{Huxley:2000aa}.}
\end{figure}

To quantify the VL rotation we define the peak splitting angle ($\omega$) shown in Fig.~\ref{DifPat}(a), determined from two-Gaussian fits to the diffraction pattern intensity.
Specific details of the fitting will be discussed in more detail later.
The temperature dependence of $\omega$ is summarized in Fig.~\ref{VLrotation}(a) for all the magnetic fields measured, together with results from our previous SANS studies obtained at base temperature~\cite{Avers:2020wx}.
At all fields the temperature dependence of $\omega$ appears to be linear within the measurement error, and extrapolate to zero well below the A-B phase transition.
The larger error bars at higher temperature is due to an increasing penetration depth and the resulting decrease in the scattered intensity~\cite{Gannon:2015ct}.

Figure~\ref{VLrotation}(b) shows $\omega$ equicontours superimposed on the {\UPt} $H$-$T$ phase diagram.
The nonmonotonic behavior, previously reported at base temperature~\cite{Avers:2020wx}, is clearly observed at higher temperatures, although with a decreasing amplitude.
Furthermore, the splitting extrapolates to zero in the zero field limit, and also decreases upon approaching the B-C phase transition.
However, once in the C phase the splitting remains at a fixed value of $\sim 8^{\circ}$~\cite{Avers:2020wx}.
At all temperatures the maximal VL rotation is observed at 0.8~T.
Also indicated in Fig.~\ref{VLrotation}(b) is the approximate temperature at 0.19~T at which $\omega$ reaches $30^{\circ}$ in the vicinity of the A phase, reported by Huxley {\em et al.}~\cite{Huxley:2000aa}.

Ensuring a reliable determination of $\omega$ requires a careful approach to the fitting.
At all fields and temperatures the radial position ($Q_R$) as well as the radial ($\Delta Q_R$) and azimuthal ($\Delta \theta$) widths were constrained to be the same for both of the split peaks.
Furthermore, the azimuthal width at each field was determined from fits at low temperature where the peaks are clearly separated, and then kept fixed at the higher temperature where they begin to overlap.
To justify this approach, we note that when the peaks are clearly separated, $\Delta \theta$ does not exhibit any systematic temperature dependence.
The azimuthal width does show a field dependence, however, with $\Delta \theta$ decreasing from $\sim 11.5^{\circ}$ FWHM at 0.4~T to $\sim 6.5^{\circ}$ FWHM at 1.2~T.

\begin{figure}
    \includegraphics[width = 67mm]{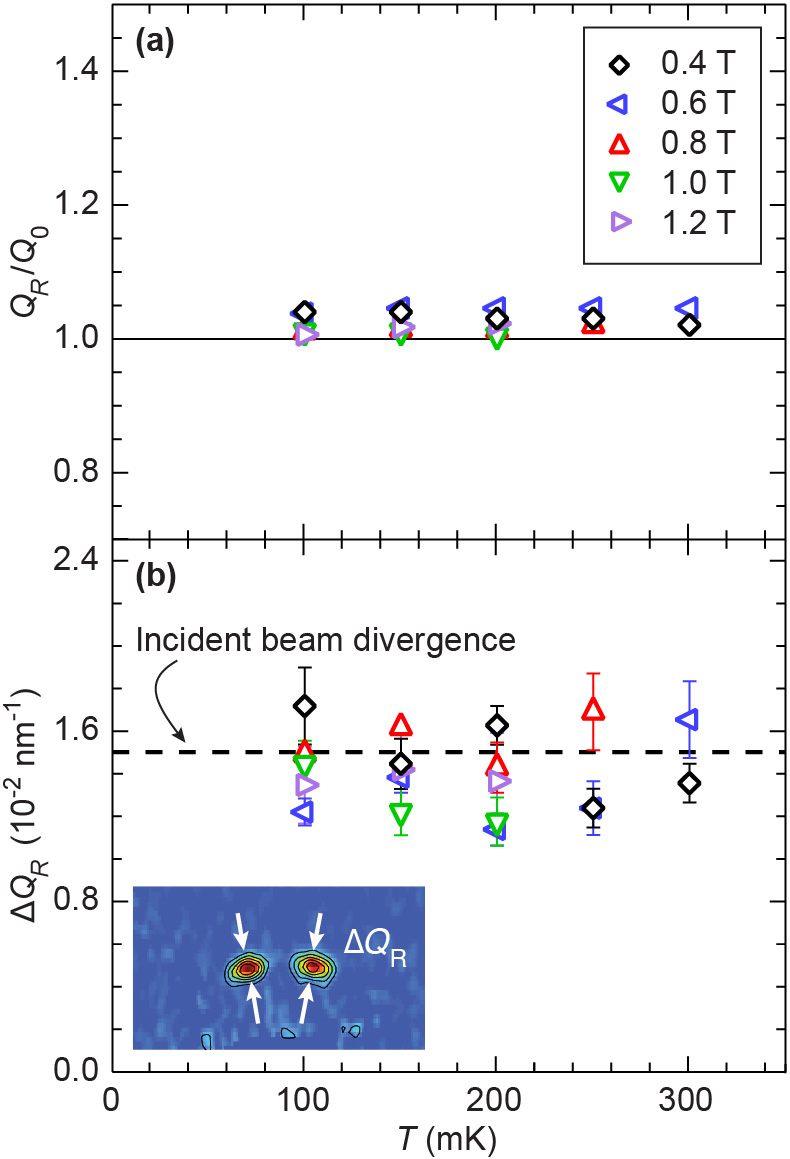}
    \caption{\label{VLdensity}
        Vortex lattice density.
        (a) Scattering vector magnitude normalized to the value expected for a triangular VL.
        (b) Radial width of the VL Bragg peaks (FWHM) compared to the incident beam divergence (dashed line).
        The inset indicates $\Delta Q_R$ within the detector plane.}
\end{figure}

The VL density is reflected in $Q_R$ and $\Delta Q_R$, shown in Fig.~\ref{VLdensity}.
The magnitude of the scattering vector in Fig.~\ref{VLdensity}(a) agrees to within a few percent with $Q_0 = 2\pi (\sqrt{3}/2)^{1/4} \sqrt{B/\Phi_0}$ expected for a triangular VL and assuming that the magnetic induction ($B$) is equal to the applied magnetic field.
Here $\Phi_0 = h/2e = 2069$~T~nm$^2$ is the flux quantum.
The small deviation between $Q_R$ and $Q_0$ is slightly greater at low fields consistent with earlier work~\cite{Avers:2020wx}, but notably independent of temperature.
Similarly, there is no systematic temperature or field dependence in the radial width in Fig.~\ref{VLdensity}(b).
However, the values are systematically at or below the divergence of the incident beam, indicating a highly ordered VL which leads to a diffracted neutron beam that is more collimated than the incident one.  

\begin{figure*}
    \includegraphics[width = 165mm]{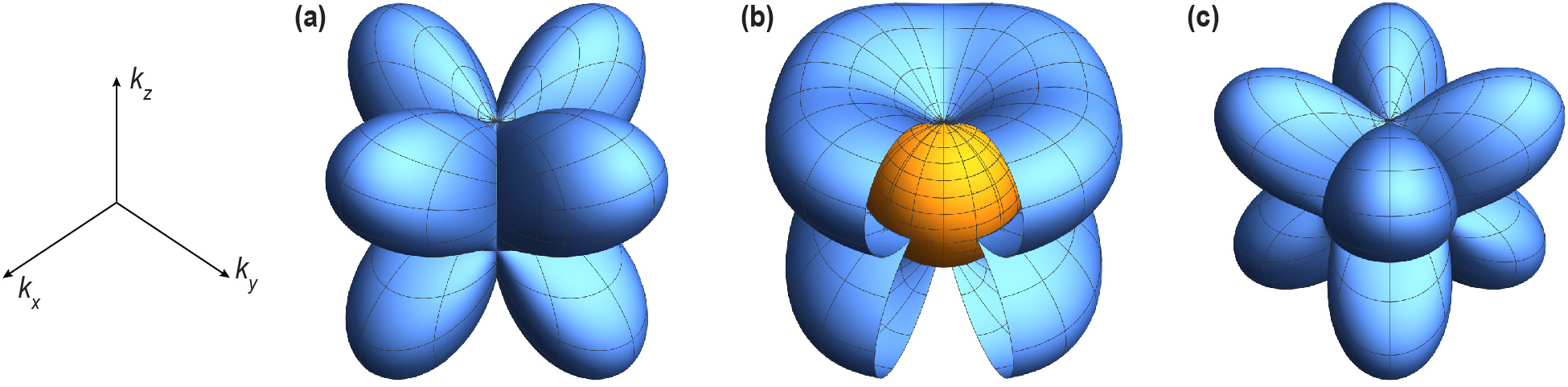}
    \caption{\label{OP}
        Order parameter in the
        (a) A phase ($\eta_1 \neq 0$, $\eta_2 = 0$),
        (b) B phase distorted by the SBF ($\eta_1 = \eta_2 = 1$, $\epsilon = 0.1$), and
        (c) C phase ($\eta_1 = 0$, $\eta_2 \neq 0$).}
\end{figure*}
%

%%%%% DISCUSSION %%%%%
\section{Discussion}
The complex VL phase diagram in Fig.~\ref{VLrotation}(b) reflects the presence of multiple competing effects.
In the following we discuss how, at the qualitative level, this phase diagram arises from the interplay between the SBF and the nodal configuration of the superconducting energy gap for the A and C phases.
A more detailed treatment of the VL structure and orientation within the A phase was provided by Champel and Mineev~\cite{Champel:2001kf}.
First, however, we note that $\omega \rightarrow 0$ in the limit $T = H = 0$.
For large vortex separations the order parameter has a vanishing effect on the VL, and the orientation with Bragg peaks along the $\textbf{a}$ axis must be due to the Fermi surface anisotropy~\cite{Huxley:2000aa,Champel:2001kf}.

In momentum space the two-component $E_{2u}$ order parameter proposed for {\UPt} is given by~\cite{Sauls:1994wd}
\begin{equation}
    \label{E2uOP}
    \Delta(\mathbf k) = \left( \eta_1 \left( k_x^2 - k_y^2 \right) \pm 2i \, \eta_2 (1 - \epsilon) \, k_x \, k_y \right) k_z.
\end{equation}
Here, $\eta_1$ and $\eta_2$ are real amplitudes which depend on temperature and magnetic field and $\epsilon$ is due to the SBF.
The A and C phases correspond to a vanishing of $\eta_2$ and $\eta_1$ respectively.
The magnitude of the SBF determines the zero-field split in the superconducting transition ($\Delta T_{\text{AB}}$) and thus the width of the A phase.
Experimentally, $\Delta T_{\text{AB}} \approx 55$~mK which yields $\epsilon \propto \tfrac{\Delta T_{\text{AB}}}{T_c} \approx 0.1$~\cite{Sauls:1994wd}.
Within the B phase both components of the order parameter are non-zero, although with different amplitudes.
Due to the SBF this imbalance persists even in the low-temperature, low-field limit where both $\eta_2$ and $\eta_1$ approach unity~\cite{Sauls:1994wd}.
The order parameter structure is illustrated in Fig.~\ref{OP}.

Within the A phase SANS studies by Huxley {\em et al.} found a VL with domains rotated by $\pm 15^{\circ}$ relative to the $\textbf{a}$ axis ($\omega = 30^{\circ}$)~\cite{Huxley:2000aa}.
The VL rotation was attributed to a competition between the sixfold Fermi surface anisotropy and the fourfold anisotropy of the nodal structure in the A phase~\cite{Huxley:2000aa,Champel:2001kf}.
Notably, the rotation persists into the B phase as indicated in Fig.~\ref{VLrotation}(b).
This is not surprising since the $\eta_1/\eta_2 \rightarrow \infty$ upon approaching the A phase from low temperature, where the B phase order parameter therefore exhibit a substantial fourfold anisotropy.
However, as $\eta_2$ increases with decreasing temperature this ratio quickly decreases, causing an abrupt transition to $\omega = 0$ around 425~mK~\cite{Huxley:2000aa}.

Due to the SBF the order parameter in the B phase preserves a degree of fourfold anisotropy, as shown in Fig.~\ref{OP}.
This anisotropy is oriented in a manner similar to the A phase, with an effect on the vortex-vortex interactions which will increase with increasing field (vortex density).
The influence of the SBF anisotropy will increase further at low temperature as the superfluid density increases~\cite{Gannon:2015ct}, even if $\epsilon$ remains fixed.
This explains the initial increase of $\omega$ with field at low temperatures, with an amplitude (0.8~T) that extrapolates to a value close to $30^{\circ}$ for $T \rightarrow 0$.

As the field is increased further and approaches the BC phase transition, $\eta_1$ decreases and finally vanish.
The C phase order parameter is rotated by $45^{\circ}$ about $k_z$ with respect to the B phase, as shown in Fig.~\ref{OP}.
This will favor a VL oriented along the $\textbf{a}$ axis, i.e. the same as the Fermi surface anisotropy, and explains the non-monotonic VL rotation as a function of field.
Once $\eta_1$ has fully vanished no further VL rotation is expected, in agreement with the observed field-independence of $\omega \approx 8^{\circ}$ in the C phase~\cite{Avers:2020wx}.

%%%%% CONCLUSION %%%%%
\section{Conclusion}
In summary, the rotated VL phase at low temperatures and intermediate fields in Fig.~\ref{VLrotation}(b) can be directly attributed to the SBF.
To our knowledge this is the first observation of such an effect at the microscopic level, and may provide further constraints on the nature of both the SBF and the order parameter in {\UPt}.
A quantitative understanding of $\omega(T,H)$ will require a detailed theoretical analysis, taking into account the field and temperature dependence of the superfluid density as well as the complex Fermi surface of {\UPt}.
Here, the finite value of $\omega$ in the C phase is somewhat surprising and not obviously consistent with the order parameter in Eq.~(\ref{E2uOP}).

%%% END MATTER %%%
\section*{Conflict of Interest Statement}
The authors declare that the research was conducted in the absence of any commercial or financial relationships that could be construed as a potential conflict of interest.

\section*{Author Contributions}
KEA, WPH, and MRE conceived of the experiment.
WJG and KEA grew and annealed the crystals.
KEA, AWDL, and MRE performed the SANS experiments with assistance from LDS.
KEA, WPH, and MRE wrote the paper with input from all authors.

\section*{Funding}
This work was supported by the the Northwestern-Fermilab Center for Applied Physics and Superconducting Technologies (KEA) and by the U.S. Department of Energy, Office of Basic Energy Sciences, under Awards No.~DE-SC0005051 (MRE: University of Notre Dame; neutron scattering) and DE-FG02-05ER46248 (WPH: Northwestern University; crystal growth and neutron scattering).
A portion of this research used resources at the High Flux Isotope Reactor, a DOE Office of Science User Facility operated by the Oak Ridge National Laboratory.

\section*{Acknowledgments}
We are grateful to J.~A.~Sauls for numerous discussions and to V. P. Mineev for valuable feed-back.

%\section*{References}
\bibliography{UPt3VLPD}

\end{document}